\documentclass[sigconf]{acmart} 

\AtBeginDocument{%
  \providecommand\BibTeX{{%
    \normalfont B\kern-0.5em{\scshape i\kern-0.25em b}\kern-0.8em\TeX}}}

\usepackage{latexsym}
\usepackage{amsmath}
\usepackage{url}
\usepackage{amsmath,nccmath}
\usepackage{algorithm}
\usepackage{algorithmic}
\usepackage{xcolor}
\usepackage{graphicx}
\usepackage{subfigure}

\usepackage{bbm}

\theoremstyle{definition}

\usepackage{enumitem}

\usepackage{array}
\newcolumntype{P}[1]{>{\centering\arraybackslash}p{#1}}

\newcommand{{\method}}{DisGNN}
\usepackage{multirow}

\acmYear{2022} 
\setcopyright{acmcopyright}\acmConference[WWW '22]{Proceedings of the ACM Web
Conference 2022}{April 25--29, 2022}{Virtual Event, Lyon, France}
\acmBooktitle{Proceedings of the ACM Web Conference 2022 (WWW '22), April
25--29, 2022, Virtual Event, Lyon, France}
\acmPrice{15.00}
\acmDOI{10.1145/3485447.3511929}
\acmISBN{978-1-4503-9096-5/22/04}

\title{Exploring Edge Disentanglement for Node Classification
}

\author{Tianxiang Zhao, Xiang Zhang, Suhang Wang}
\affiliation{%
   \institution{College of Information Sciences and Technology, The Pennsylvania State University}
   \city{State College}
   \state{PA}
   \country{USA}}
\email{{tkz5084,xzz89,szw494}@psu.edu}

\begin{document}
\begin{abstract}
Edges in real-world graphs are typically formed by a variety of factors and  carry diverse relation semantics. For example, connections in a social network could indicate friendship, being colleagues, or living in the same neighborhood. However, these latent factors are usually concealed behind mere edge existence due to the data collection and graph formation processes. Despite rapid developments in graph learning over these years, most models take a holistic approach and treat all edges as equal. One major difficulty in disentangling edges is the lack of explicit supervisions. In this work, with close examination of edge patterns, we propose three heuristics and design three corresponding pretext tasks to guide the automatic edge disentanglement. Concretely, these self-supervision tasks are enforced on a designed edge disentanglement module to be trained jointly with the downstream node classification task to encourage automatic edge disentanglement. Channels of the disentanglement module are expected to capture distinguishable relations and neighborhood interactions, and outputs from them are aggregated as node representations. The proposed {\method} is easy to be incorporated with various neural architectures, and we conduct experiments on $6$ real-world datasets. Empirical results show that it can achieve significant performance gains.

\end{abstract}



\maketitle

\section{Introduction}

In recent years, learning from graph-structured data is receiving a growing amount of attention due to the ubiquity of this data form in the world. For example, social networks~\cite{Fan2019GraphNN,Zhong2020MultipleAspectAG}, molecular structures~\cite{Mansimov2019MolecularGP,Chereda2019UtilizingMN} and knowledge graphs~\cite{Sorokin2018ModelingSW} all require utilizing the rich connection information among nodes. Graph neural networks (GNNs)~\cite{wu2020comprehensive} provide a general and effective framework for mining from graphs, and is developing rapidly among the years. 

\begin{figure}[t!]
  \centering
		\includegraphics[width=0.42\textwidth]{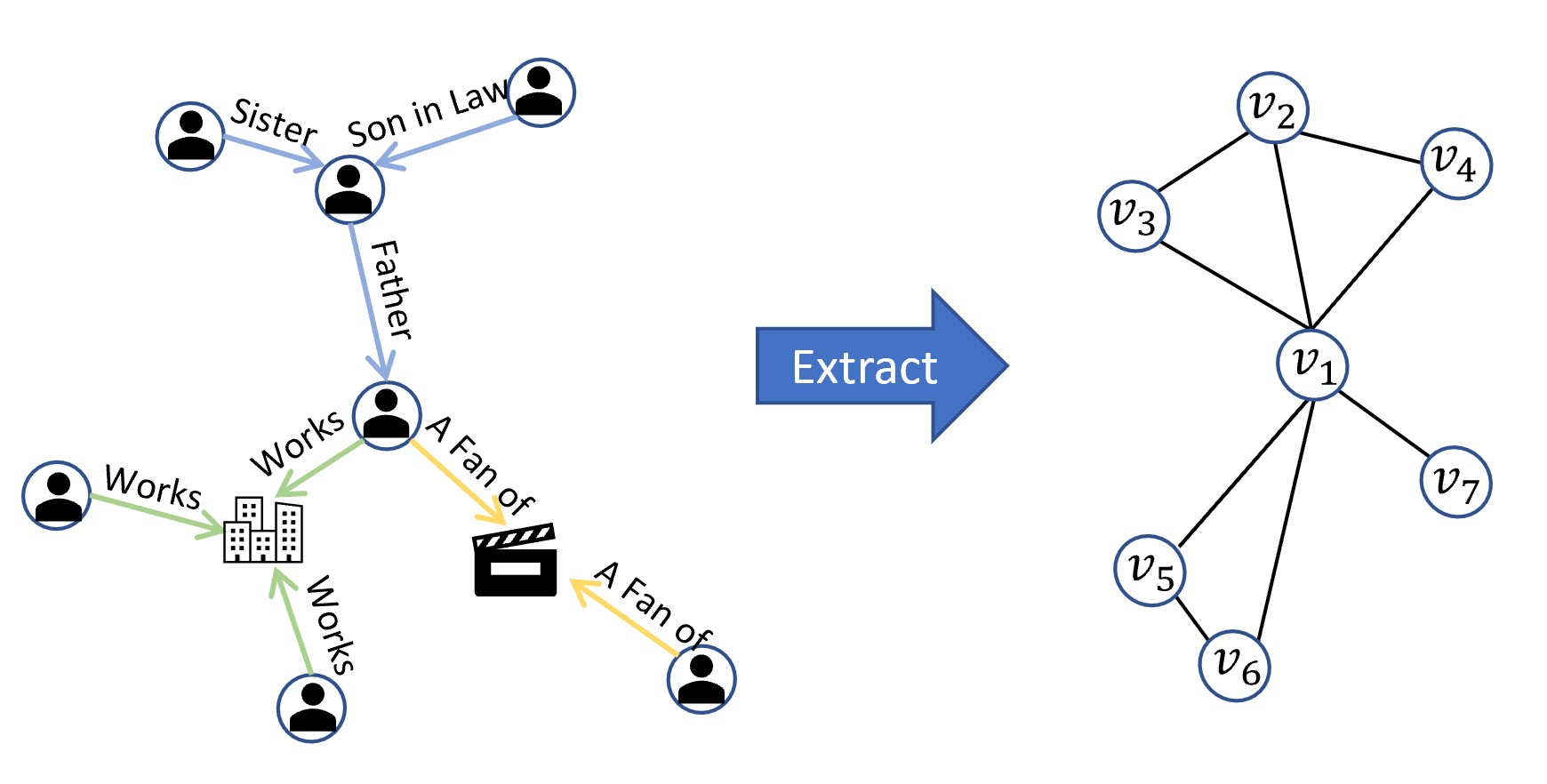}
    \vskip -2em
    \caption{An example on Facebook. There are multiple possible relations behind edge existence, but such information is usually latent and unavailable during graph formation. } \label{fig:example}
  \setlength{\abovecaptionskip}{0cm}
\end{figure}

GNNs can be viewed as a message-passing network~\cite{gilmer2017neural}, which conduct node representation updating and neighborhood interaction modeling iteratively. Variants of GNNs~\cite{Kipf2017SemiSupervisedCW,Hamilton2017InductiveRL,xu2018powerful} mainly differ in the proposed mechanism of aggregating messages from node neighborhoods. However, despite improved representation power from architecture designs, most of them adopt a uniform processing on edges in the input graph. Their model is designed under the assumption that all edges are generated under the same latent distribution and indicate the same relationship, hence those edges are taken as equal. Yet real-world graphs are usually formed via a complex and heterogeneous process, with multiple possible causes behind existence of edges. For example, as shown in Figure~\ref{fig:example}, connection in a social network could be resulted from similarity in interest, colleague relationship, or living in the same neighborhood, etc. Neglect of semantic differences carried by edges would harm the modeling of neighborhood interactions and the quality of obtained node embeddings. This observation motivates us to utilize each edge with considerations on its carried semantic relationship. 

However, incorporating edge disentanglement into graph learning is a difficult task. The major problem is the lack of supervision signals to guide the learning process. Due to limitations in data collection and graph formation procedures, most graph datasets have plain edges without attributes, making it difficult to obtain disentangled edge set. There are few pioneering works~\cite{ma2019disentangled,yang2020factorizable} in automatic disentanglement of graphs. However, ~\citet{ma2019disentangled} focuses on obtaining node-level representations w.r.t disentangled latent factors and pays little attention to latent multi-relations of edges, while ~\citet{yang2020factorizable} mainly focuses on graph-level tasks and proposes to disentangle the input graph into multiple factorized sub-graphs. Furthermore, none of these methods consider the problem of providing extra supervision signals for edge disentanglement, and rely upon the holistic learning process instead.

Targeting at this problem, in this work, we focus on facilitating graph learning, particularly the node classification task, with automatic edge disentanglement. We propose an edge disentanglement module, and make the first attempt in designing a set of pretext tasks to guide the edge disentanglement. It is a difficult task, because the latent multi-relations behind edge existence are usually hidden and unavailable in real-world datasets. With a close examination of heuristics, we propose three self-supervision signals: (1) The union of disentangled edges should recover the original input edge set, (2) Edges carrying homophily information (homo-edges) models intra-class interactions while edges carrying heterophily property (hetero-edges) models inter-class interactions, hence they usually carry different semantics, (3) The edge patterns and interactions captured by different channels of {\method} (each represents a disentangled relation group) should be discrepant, with a clear difference in distribution. 

Concretely, the designed edge disentanglement module is incorporated into existing GNN layers, and three self-supervision tasks corresponding to aforementioned heuristics are implemented and applied to them. This approach, {\method}, enables separate processing of disentangled edges during learning towards node classification via optimizing the disentanglement-eliciting pretext tasks jointly. The proposed method is evaluated on six real-world graphs, and we further conduct ablation studies to analyze its behaviors. Experimental results show that the proposed approach is effective and promising, achieving improvements on all datasets.

To summarize, in this work, we studies node classification with automatic edge disentanglement and make following contributions: 
\begin{itemize}
    \item \textbf{Node classification in the edge-disentanglement-aware manner.} Usually there are multiple latent factors behind the existence of edges between node pairs, and utilizing them could improve node embedding quality and help the node classification task. In this work, we study this problem by designing and applying an edge disentanglement module in together with GNN layers.
    \item \textbf{Automatic edge disentanglement with pretext tasks.} Although edges in real-world graphs usually contain a variety of different relations in nature, lack of supervision limits the discovering of them. In this work, we propose three self-supervision signals, which could encourage our designed module to automatically disentangling edges.
    \item \textbf{Evaluation on real-world graph sets.} We evaluate the proposed approach on six real-world datasets, and results show the improvement brought by the proposed approach.
\end{itemize}

\section{Related Work}

\subsection{Graph Neural Network}
Graph neural networks (GNNs) are developing rapidly in recent years, with the increasing needs of learning on relational data structures~\cite{Fan2019GraphNN,dai2022towards,zhao2021graphsmote,zhao2020semi}. Generally, 
existing GNNs can be categorized into two categorizes, i.e., spectral-based approaches~\cite{bruna2013spectral,Tang2019ChebNetEA,Kipf2017SemiSupervisedCW} based on graph signal processing theory, and spatial-based approaches~\cite{duvenaud2015convolutional,atwood2016diffusion,xiao2021learning} relying upon neighborhood aggregation. Despite their differences, most GNN variants can be summarized with the message-passing framework, which is composed of pattern extraction and interaction modeling within each layer~\cite{gilmer2017neural}. 

Dedications have been made towards mining richer information from the provided relation topology. For example, ~\citet{Velickovic2018GraphAN} extends self-attention mechanism to enable learning the weights for nodes inside the neighborhood. ~\citet{xu2018powerful} extends expressive power of GNNs to the same order of WL test by designing an injective neighborhood aggregator. ~\citet{abu2019mixhop} extracts multi-hop neighborhoods and learns to mix them for improving center node representations. ~\citet{dai2021nrgnn} studies noisy information propagating along given edges and proposes to update network structure to improve robustness. However, all these methods are holistic approaches and neglect the latent factors behind edge existence. In real world cases, connections among entities are usually multi-relational in nature. Although they are given in the form of plain edges, different edges could be caused by different factors and represent different semantics. Disentangling these factors could enable us to utilize given edges with the awareness of their carried latent relations, and model richer node interactions.

Discovering latent factors behind graphs and utilizing them to improve graph learning is an important but under-exploited problem.  Currently, works in this direction are rather limited. One major difficulty of this task is the lack of explicit supervisions on ground-truth disentanglement. For example, graph attention network (GAT)~\cite{Velickovic2018GraphAN} offers a mechanism of specifying different weights to nodes in the neighborhood, and has the potential of disentangling relation types with multiple attention heads. However, analysis find that GAT tends to learn a restricted ``static'' form of attention and the patterns captured by different heads are limited~\cite{brody2021attentive,kim2020find}. A pretext task with self-supervision on attention heads is designed in ~\cite{kim2020find}, but it only encourages recovering ground-truth edges as a whole and provides no signals on relation disentanglement. DisenGCN~\cite{ma2019disentangled} makes the first attempt by learning node embeddings with respect to different factors with a neighborhood routing mechanism. IPGDN~\cite{liu2020independence} further improves on that by utilizing Hilbert-Schmidt Independence Criterion to promote independence among disentangled embeddings. FactorGCN~\cite{yang2020factorizable} studies the disentanglement of graphs into multiple sub-graphs mainly for graph-level tasks. Different from these methods, this work focuses on automatic disentanglement at the edge level, and designs three pretext tasks providing heuristics to guide the training process.

\subsection{Self-supervision in Graph}
Self-supervised learning targets at extracting informative knowledge through well-designed pretext tasks without relying on manual labels, and is able to utilize large amount of available unlabeled data samples. This framework has been shown to be promising in eliminating heavy label reliance and poor generalization performance of modern deep models. Besides its success in computer vision~\cite{jing2020self} and natural language processing~\cite{lan2019albert}, there is a trend of developing this idea in the graph learning domain~\cite{liu2021graph,hu2019strategies}.

Rich node and edge features encoded in the graph are promising in providing self-supervision training signals to guide the graph learning process. Pretext tasks exploiting multiple different levels of graph topology have been designed to boost the performance of graph learning models. Existing approaches can be categorized into three types based on the designed pretext signals, including node-level signals~\cite{hu2019strategies}, edge-level signals~\cite{kim2020find} and graph-level signals~\cite{zeng2020contrastive}. ~\citet{hu2019strategies} captures local node similarity via randomly masking nodes and learning to recover them from neighbors. ~\cite{jin2020self} predicts pairwise node distance on graphs to encode global-structure information. ~\citet{qiu2020gcc} uses contrastive learning on randomly sampled sub-graphs to highlight local-neighborhood similarities. In the classical semi-supervised setting on graphs, which contains a vast number of unlabeled nodes, these auxiliary tasks help to make learned representations more general, robust to noises, and more transferable~\cite{zeng2020contrastive}. A more complete survey of existing self-supervised learning on graphs can be found in ~\cite{liu2021graph,xie2021self} 

Unlike the aforementioned approaches, we target at designing pretext signals to guide the learning of attention distributions within the neighborhood.  \citeauthor{kim2020find}~\cite{kim2020find} propose to supervise graph attention heads with real input edges, encouraging higher attention score between connected nodes while lower score between unconnected pairs. However, they aim to remove noises in edges and apply the same signal to all attention heads in the same manner. In contrast, we focus on disentangling relations and encouraging different attention heads to capture different semantics.

\section{Problem Formulation}

We use $\mathcal{G} = \{\mathcal{V}, \mathbf{A}, \mathbf{F}\}$ to denote an attributed network, where $\mathcal{V}=\{v_1,\dots,v_{n}\}$ is a set of $n$ nodes. $\mathbf{A} \in \mathbb{R}^{n \times n}$ is the adjacency matrix of $\mathcal{G}$. $\mathbf{F} \in \mathbb{R}^{n \times d}$ denotes the node attribute matrix, where $\mathbf{F}[j,:] \in \mathbb{R}^{1 \times d}$ is the $d$-dimensional node attributes of node $v_j$. In real-world, each edge is formed due to various reasons and carries rich semantic meanings. For example, in Facebook, $v_i$ and $v_j$ are linked because they are ``collegaes''; $v_i$ and $v_k$ are connected because they are both interested in ``Football''. Hence, in learning the features of $v_i$, the utilization of $v_j$ and $v_k$ should be different in a relation-aware manner. However, for many real-world graphs such as Facebook, we only know that there exists such factors but the factors of each edge is not explicitly known. Thus, in this paper, we assume that we are only given $\mathcal{G} = \{\mathcal{V}, \mathbf{A}, \mathbf{F}\}$ without edge attributes and aims to disentangle the graph to facilitate node classification. 

As shown in ~\cite{gilmer2017neural}, most existing GNN layers can be summarized in the following equations:
\begin{equation}
    \begin{aligned}
    \mathbf{m}_{v}^{l+1} &=\sum_{u \in \mathcal{N}(v)} M_{l}\large( \mathbf{h}_v^l, \mathbf{h}_u^l, \mathbf{A}_{v,u} \large) \\
    \mathbf{h}_v^{l+1} &= U_{l}\large(\mathbf{h}_v^{l}, \mathbf{m}_{v}^{l+1} \large)
    \end{aligned}
\end{equation}
where $\mathcal{N}(v)$ is the set of neighbor of $v$ in $\mathcal{G}$ and $\mathbf{h}_{v}^{l}$ denotes representation of $v$ in the $l$-th layer. $\mathbf{A}_{v,u}$ represents the edge between $v$ and $u$. $M_{l}$, $U_l$ are the message function and update function respectively at layer $l$. However, this framework neglects diverse relations behind edges. With a disentangled adjacency matrix set $\mathcal{A}^{dis} = \{ \mathbf{A}^0, \mathbf{A}^1, ...  \mathbf{A}^m \}$, in which $\mathbf{A}^i \in \mathbb{R}^{n \times n}$ and $\mathbf{A}^i \subset \mathbf{A}$, we can obtain edge sets representing different relation semantics. Then, the GNN layer could be be changed into:
\begin{equation}
    \begin{aligned}
    \mathbf{m}_{v}^{l+1, i} &=\sum_{u \in \mathcal{N}_i(v)} M_{l}^i\large( \mathbf{h}_v^l, \mathbf{h}_u^l, \mathbf{A}_{v,u}^i \large) \\
    \mathbf{h}_v^{l+1} &= U_{l}\large(\mathbf{h}_v^{l+1}, \mathbf{m}_{v}^{l+1,0}, \dots, \mathbf{m}_{v}^{l+1,m} \large)
    \end{aligned}
    \label{eq:disGNN}
\end{equation}
where $\mathcal{N}_i(v)$ is the set of neighbors of $v$ in adjacency matrix $\mathbf{A}^i$. This disentangled GNN enables heterogeneous processing of given edges, and may produce more discriminant node embeddings for downstream tasks.

In this work, we focus on semi-supervised node classification task. $\mathbf{Y} \in \mathbb{R}^{n}$ is the class information for nodes in $\mathcal{G}$. During training, only a subset of $\mathbf{Y}$ is available, and the task is to train a classifier for unlabeled nodes. We use $\mathcal{V}_L$ and $\mathbf{Y}_L$ to represent the set of supervised nodes and their corresponding labels respectively.

\vspace{0.5em}
\noindent{}\textit{Given $\mathcal{G}= \{\mathcal{V}, \mathbf{A}, \mathbf{F}\}$, and labels $\mathbf{Y}_{L}$ for a subset of nodes $\mathcal{V}_{L}$, we aim to learn a node classifier $f$. $f$ should learn to disentangle relations behind edge existence as $\mathcal{A}^{dis}$ and model node interactions accordingly during the label prediction process:
\begin{equation}
    f(\mathcal{V}, \mathbf{A}, \mathbf{F}) \rightarrow \mathbf{Y}
\end{equation}
}

\section{Methodology}\label{sec:overview}

In this work, we implement the model as composing of two parts: a GNN-based feature extractor for obtaining node representations and a MLP-based classifier for node classification. $\theta, \xi$ are used to represent parameters of these two components respectively. The details of them will be introduced below. 

\subsection{Model Architecture}
\subsubsection{Feature Extractor}
\begin{figure}[t!]
  \centering
    \includegraphics[width=0.5\textwidth]{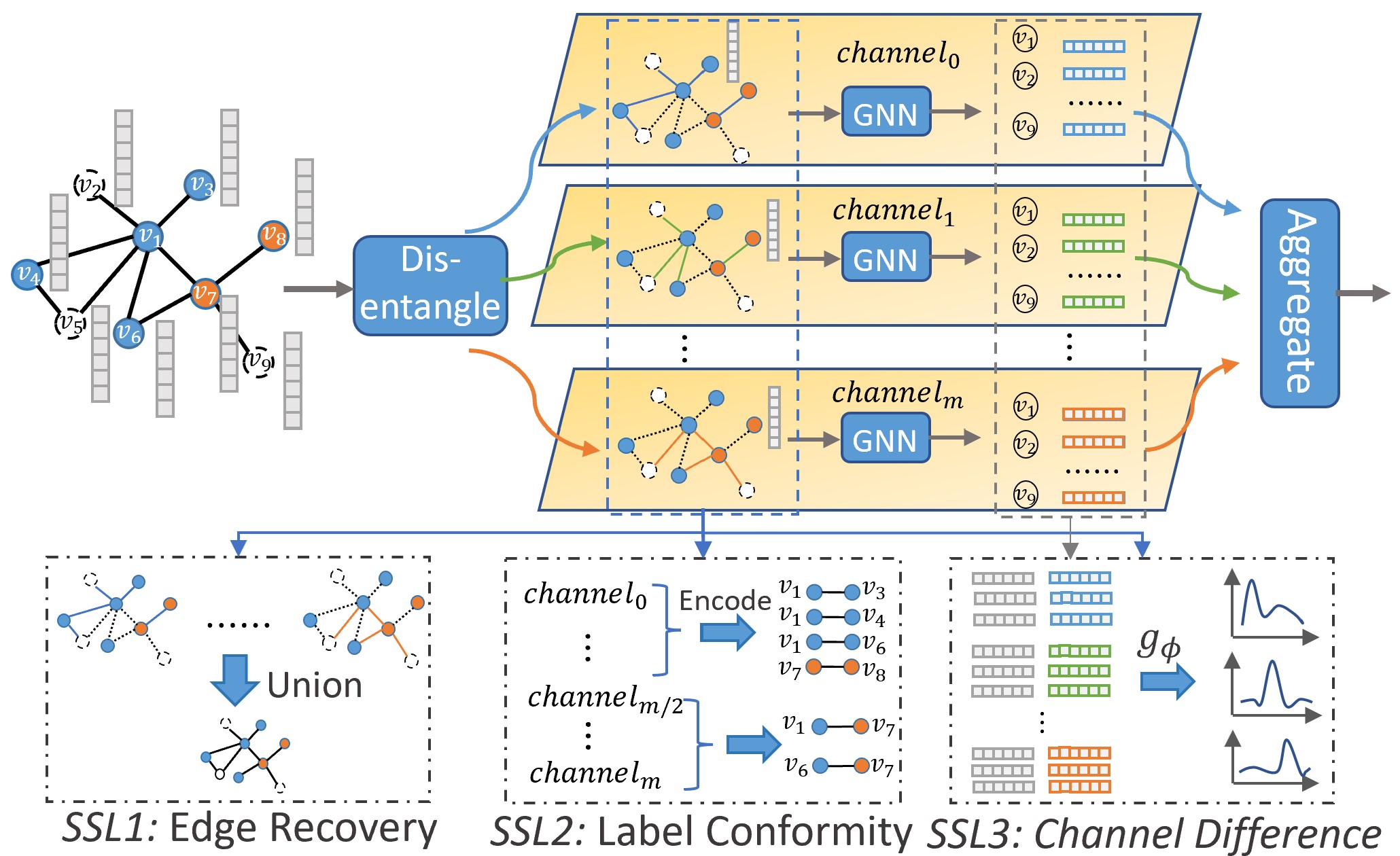}
    \vskip -1em
    \caption{Overview of the {\method} layer.} \label{fig:overview}
  \setlength{\abovecaptionskip}{0cm}
\end{figure}

The feature extractor is adopted to encode graph structures and obtain node embeddings for classification. Following existing works~\cite{zhu2020beyond,zhao2021graphsmote}, it can be implemented by stacking two GNN layers. In this work, we design and utilize two {\method} layers. Each {\method} layer is implemented with the edge disentanglement module and GNN variants following Equation~\ref{eq:disGNN}, upon which we apply our proposed self-supervision tasks. An overview of the proposed {\method} layer is shown in Figure~\ref{fig:overview}. The feature extractor is optimized with gradients from both node classification and self-supervision tasks, and its details will be presented in the next section. The parameter set of this extractor is denoted as $\theta$.

\subsubsection{Classifier}
Based on the extracted node representation $\mathbf{h}_{v}$ for node $v$ by feature extractor, a classifier is trained to perform node classification. Specifically, we use a $2$-layer MLP:
\begin{equation}
    \begin{aligned}
    \mathbf{P}_{v}  = softmax \big( \mathbf{W}_{2}^{cls} \cdot \sigma(\mathbf{W}_1^{cls} \cdot \mathbf{h}_{v}) \big),
    \end{aligned}
\end{equation}
where $\mathbf{W}_{1}^{cls},\mathbf{W}_{2}^{cls}$ are the parameter matrices of this classifier head. And the training objective $\mathcal{L}_{node}$ on this task would be:
\begin{equation}
  \min_{\theta,\xi} -\sum_{v \in \mathcal{V}_{sup}} \sum_{c \in \mid Y \mid} Y_{v}^c \cdot log(\mathbf{P}_{v}^c).
  \label{eq:node_cls}
\end{equation}
$\xi$ denotes parameters in this classification head, containing $\mathbf{W}_{1}^{cls}$ and $\mathbf{W}_{2}^{cls}$. $\mathcal{V}_{sup}$ represents supervised nodes during training, $Y_v^c$ indicates whether node $v$ belongs to class $c$, $|Y|$ represents the set of node classes, and $\mathbf{P}_{v}^c$ is the predicted probability of $v$ in class $c$.

\subsection{{\method} Layer}
\subsubsection{Edge Disentanglement Module}


With the input of graph $\mathcal{G}$, this module is designed to obtain a set of disentangled edge distributions, $\mathcal{A}^{dis}$. As edges are unattributed and binary in most graphs, we analyze the edge property through comparing the node pair connected by it. One module with $m+1$ channels can be constructed, with each channel capturing the dynamics reflecting one different group of relations. Each relation-modeling channel can be constructed in various ways, like dot-production, linear layer, neural network, etc.

In this work, to keep both simplicity and representation ability, we implement each relation-modeling channel as a MLP. Concretely, inside the $l$-th {\method} layer, predicted edge between node pair $v$ and $u$ generated by the $i$-th channel can be written as:

\begin{equation}
    \begin{aligned}
    \hat{e}_{v,u}^{l,i} &= sigmoid \big( \mathbf{W}_2^{l,i} LeakyReLU \big( \mathbf{W}_{1}^{l,i} \cdot [\mathbf{h}^{l}_{v}, \mathbf{h}^{l}_{u} ] \big) \big) \\
    \mathbf{A}_{v,u}^{l, i} &= \frac{exp({\hat{e}_{v,u}^{l,i}})}{\sum_{w \in \mathcal{V}}exp({\hat{e}_{v,w}^{l,i}})}
    \end{aligned}
    \label{eq:used_at}
\end{equation}
In this equation, $\mathbf{h}_{u}^{l}$ corresponds to the extracted embedding of node $u$ as input to layer $l$, and  $\mathbf{h}_{u}^{l},\mathbf{h}_{v}^{l}$ are concatenated as input. $\mathbf{W}_{1}^{l,i},\mathbf{W}_{2}^{l,i}$ are the parameter matrices, and $\hat{e}_{v,u}^{l,i}$ is the edge existence probability between $v$ and $u$ modeled by channel $i$ in the $l$-th layer. $\hat{e}$ is only calculated for connected node pairs to reduce computation complexity and avoid introducing noises. Following the pre-processing strategy in ~\cite{Kipf2017SemiSupervisedCW}, edge weights in $\mathbf{A}^{l,i}$ are normalized before using as input into GNN layers.

\subsubsection{Incorporating into GNN Layer}

The edge disentanglement module is easy to be incorporated into various modern GNN layers through taking each disentangled edge group as a channel for modeling neighborhood interactions following Equation~\ref{eq:disGNN}. In this subsection, we take the GNN layer proposed in ~\cite{Velickovic2018GraphAN} as an example. The process of update node representations on each channel can be written as:
\begin{equation}\label{eq:GNN_main}
\begin{aligned}
\mathbf{h}_{v}^{l+1,i} = \sum_{u \in \mathcal{V}} \mathbf{A}_{u,v}^{l,i} \cdot \mathbf{W}_{feat}^{l,i}\mathbf{h}_{u}^{l},
\end{aligned}
\end{equation}
in which $\mathbf{W}_{feat}^{l,i}$ is the parameter matrix for learning embedding features of $i$-th channel in the $l$-th layer.

Then, we aggregate representation learned on each channel via concatenation and a $1$-layer neural network:
\begin{equation}
    \begin{aligned}
    \mathbf{h}_{v}^{l+1} = \sigma \big( \mathbf{W}_{agg}^{l}( [ \mathbf{h}^{l+1,0}_{v},  \mathbf{h}^{l+1,1}_{v}, \dots, \mathbf{h}^{l+1,m}_{v}]) \big).
    \end{aligned}
\end{equation}
$\sigma$ represents nonlinear activation function, and we implement it as LeakyRelu. $\mathbf{W}_{agg}^{l}$ is a weight matrix used to fuse representations learned by each channel. Note that it is not limited to using this specific GNN variant, and we further test it on other typical GNN layers in the experiments.

\subsection{Self-supervision for Disentanglement}
One major difficulty in disentangling factors behind edges is the lack of supervision signals. Although it is often the case that edges contain multiple different relations and carry different semantics in a graph, it is difficult to collect ground-truth edge disentanglement as the supervision, as the example in Figure~\ref{fig:example}. Hence, in this part, we introduce how we manage to discover them automatically with carefully-designed self-supervisions, which also makes up the main contribution of this work. An overview is provided in Figure~\ref{fig:overview}.

\subsubsection{Signal $1$: Edge Recovery} 
One most direct way of providing prior knowledge to guide the learning of relation distribution is utilizing the input edges. If node $u$ and $v$ are linked, then certain relationship exists between them, and such relation should be captured by the edge disentanglement module. Note that channels of it are expected to capture different relations, hence we require the union of them to successfully recover given edges. Specifically, based on the nonlinear property of \textit{sigmoid} activation function, we propose to implement this pretext task on disentanglement channels via:
\begin{equation}
    \begin{aligned}
    \bar{e}_{v,u}^{l} &= sigmoid \big( \sum_{i=0}^m \mathbf{W}_2^{l,i} LeakyReLU \big( \mathbf{W}_{1}^{l,i} \cdot [\mathbf{h}^{l}_{v}, \mathbf{h}^{l}_{u} ] \big) \big) 
    \end{aligned}
\end{equation}
$\bar{e}_{v,u}^{l}$ represents the union of predicted edges between $u$ and $v$ at layer $l$, and supervision can be applied on top of it. Usually, the number of nodes is large with high sparsity degrees, hence it is not efficient to use positive and negative node pairs all at once. Addressing it, we sample a batch of positive node pairs $E_p \in \{(v,u) \mid \mathbf{A}_{v,u} \neq 0 \} $ and negative node pairs  $E_n \in \{(v,u) \mid \mathbf{A}_{v,u} = 0 \} $ each time as the supervision. The size of $E_p$ is set as $p_e \cdot | \mathcal{A}^+ |$, with $p_e$ and $ \mathcal{A}^+$ being the sampling ratio and the set of connected node pairs respectively. The size of $E_n$ is set as  $\min(p_e \cdot  |\mathcal{A}^-|, 3\cdot | E_p|)$, and $\mathcal{A}^-$ represents the set of unconnected node pairs. By default, $p_e$ is set as $1.0$.

Specifically, the self-supervision loss $\mathcal{L}_{edge}$ is implemented as:
\begin{equation} \small
    \begin{aligned}
    \min_{\theta} \mathcal{L}_{edge} =  - \frac{1}{| E_p \cup E_n |} \sum_l  \Big( \sum_{(v,u)\in  E_p} \log \bar{e}_{v,u}^{l} + \sum_{(v,u)\in  E_n} \log (1-\bar{e}_{v,u}^{l})\Big)
    \end{aligned}
    \label{eq:edge_recovery}
\end{equation}

\subsubsection{Signal $2$: Label Conformity}  Another heuristics we propose is analyzing the conformity of labels between two nodes connected by edges. Due to the lack of explicit information, it is not straightforward to design direct supervisions on edge disentanglement. However, following observations in ~\cite{zhu2020beyond}, we notice that given edges can be split into two groups: homo-edges modeling intra-class interactions and hetero-edges modeling inter-class interactions. These two groups usually carry different relations and can serve as a heuristic for automatic disentanglement. Basing on this observation, we can obtain two pseudo edge sets:
\begin{equation}
    \begin{aligned}
    E^{homo} = \{ \mathbf{A}_{v,u} \mid \mathbf{A}_{v,u} \neq 0, Y_{u}=Y_{v} \}\\
    E^{hetero} = \{ \mathbf{A}_{v,u} \mid \mathbf{A}_{v,u} \neq 0, Y_{u} \neq Y_{v} \}\\
    \end{aligned}
\end{equation}
As it is difficult to explicitly define the correspondence between edge sets and each disentanglement channel, we adopt a soft assignment, assuming that the union of half channels capture an edge set: 
\begin{equation}
    \begin{aligned}
    \bar{e}_{v,u}^{l,homo} = sigmoid \big( \sum_{i=0}^{m/2} \mathbf{W}_2^{l,i} LeakyReLU \big( \mathbf{W}_{1}^{l,i} \cdot [\mathbf{h}^{l}_{v} || \mathbf{h}^{l}_{u} ] \big) \big)
    \end{aligned}
\end{equation}

\begin{equation}
    \begin{aligned}
    \bar{e}_{v,u}^{l,hetero} = sigmoid \big( \sum_{i=m/2+1}^{m} \mathbf{W}_2^{l,i} LeakyReLU \big( \mathbf{W}_{1}^{l,i} \cdot [\mathbf{h}^{l}_{v} || \mathbf{h}^{l}_{u} ] \big) \big)
    \end{aligned}
\end{equation}

$\bar{e}_{v,u}^{l, homo},\bar{e}_{v,u}^{l, hetero}$ are expected to capture homo-edges and heter-edges respectively at layer $l$, and we supervise them in the same manner as Equation ~\ref{eq:edge_recovery}. With sampled positive set $E_p^{homo} \in E^{homo}$ and negative set $E_n^{homo} \in \mathbf{A}\setminus E^{homo} $,  $\mathcal{L}_{homo}$ is implemented as:
\begin{equation}
    \begin{aligned}
    \min_{\theta} - \frac{1}{| E_p^{homo} \cup E_n^{homo} |} &\sum_{(v,u)\in  E_p^{homo} \cup E_n^{homo} } \sum_l \large( \mathbf{1}((v,u) \in E_p^{homo}) \\ \cdot log \bar{e}_{v,u}^{l,homo} &+ \mathbf{1}((v,u) \in E_n^{homo}) \cdot log (1-\bar{e}_{v,u}^{l,homo}) \large) ,
    \end{aligned}
    \label{eq:edge_recovery}
\end{equation}
similarly is $\mathcal{L}_{hetero}$. The label conformity loss $\mathcal{L}_{conform}$ is implemented as:
\begin{equation}
\mathcal{L}_{conform} = \mathcal{L}_{homo} + \mathcal{L}_{hetero}    
\end{equation}

In experiments, we found that the amount of homo-edges and hetero-edges identified with given supervisions are sufficient for training. In cases when supervision is limited and cannot obtain a sufficient large set of labeled edges, a classifier can be trained to give them pseudo labels.

\subsubsection{Signal $3$: Channel Difference}
The last signal we designed is to promote the difference in captured patterns across channels in edge disentanglement module, as relations embedded in well-disentangled edges should be distinguishable from each other. However, directly maximizing the difference of learned knowledge by each channel is a non-trivial task. Consider that as channels are expected to model different types of relations, the semantic information learned by them would also be different, which can be reflected by comparing node embedding changes produced by them. With this observation, we first obtain channel-wise node representations:
\begin{equation}
    \begin{aligned}
    \tilde{\mathbf{h}}_{v,i}^{l} = Concat(\mathbf{h}_{v}^{l-1}, \mathbf{h}_{v}^{l, i}) 
    \end{aligned}
\end{equation}
In the following part, we will omit layer number $l$ and replace $\tilde{\mathbf{h}}_{v,i}^{l}$ with $\tilde{\mathbf{h}}_{v,i}$ for simplicity. $\tilde{\mathbf{h}}_{v,i}$ contains the semantics modeled by channel $i$. Then, we encourage distinguishability of these channel-wise representations by giving them pseudo labels w.r.t the channel.

Concretely, a label  $\tilde{{Y}}_{v,i}$ is given to each $\tilde{\mathbf{h}}_{v,i}$, which is set as the index of channel $\tilde{{Y}}_{v,i}=i$. Then, a channel discriminator $g$ is adopted to optimize this classification task. $g$ is parameterized with $\phi$:
\begin{equation}
    \begin{aligned}
    \tilde{\mathbf{P}}_{v,i}  = softmax \big( g_{\phi}( \tilde{\mathbf{h}}_{v,i} ) \big),
    \end{aligned}
\end{equation}
\begin{equation}
  \min_{\theta, \phi}\mathcal{L}_{channel} = -\sum_{v \in \mathcal{V}} \sum_{i=0}^m\sum_{c=0}^m  \mathbf{1}(\tilde{{Y}}_{v,i}=c)\log(\tilde{\mathbf{P}}_{v,i} [\tilde{Y}_{v,i}] ).
  \label{eq:diversity}
\end{equation}
$\tilde{\mathbf{P}}_{v,i} \in \mathbb{R}^{m+1}$ denotes the predicted class distribution for embedding $\tilde{\mathbf{h}}_{v,i}$, and $\mathcal{L}_{channel}$ is a cross entropy loss for embedding classification.  $g$ is implemented as a $2$-layer MLP, and we train it end-to-end with feature extractor parameterized by $\theta$. This task requires each channel to encode distinguishable semantics as well as being discriminative, hence is helpful for disentanglement.

\subsection{Optimization}
In experiments, these self-supervisions are applied to all layers in feature extractor, and are trained in together with the node classification task. Putting everything together, the full training objective is implemented as follows:

\begin{equation}
    \mathcal{L}_{full} = \mathcal{L}_{node} + \lambda_1 \mathcal{L}_{edge} +\lambda_2\mathcal{L}_{conform}+ \lambda_3\mathcal{L}_{channel},
\end{equation}
in which $\lambda_1,\lambda_2,\lambda_3$ are weight of each pretext task respectively. 

During training, we find that it would be more stable to alternatively optimize on this augmented objective and the vanilla node classification task. It can be seen as asynchronously adapt classifier to changes in edge disentanglement. We set the number of alternating steps as $n\_step$ and the full optimization algorithm is summarized in Algorithm~\ref{alg:Framwork}. From line $2$ to line $4$, we perform conventional node classification to fine-tune the model. From line $5$ to line $6$ we process input for conducting pretext tasks. And the full loss is calculated and optimized from line $7$ to line $10$. In experiments, $n\_step$ is fixed as $5$.

\begin{algorithm}[t]
  \caption{Full Training Algorithm}
  \label{alg:Framwork}
  \begin{algorithmic}[1] 
  \REQUIRE 
    $\mathcal{G} = \{\mathcal{V}, \mathbf{A}, \mathbf{F}\}$, node labels $\mathbf{Y}$, self-supervision weights $\lambda_1,\lambda_2,\lambda_3$, initial model parameter $\theta,\xi, \phi$, alternating step $n\_step$
    \WHILE {Not Converged}
    \FOR{$step$ in $ n\_step$}
    \STATE Update $\theta,\xi$ with gradients to minimize Loss $\mathcal{L}_{node}$;
    \ENDFOR
    \STATE Input $\mathcal{G}$ to the feature extractor and classifier;
    \STATE Use sampling to get training instances for self-supervision tasks;
    \STATE Calculating $\mathcal{L}_{edge}, \mathcal{L}_{conform}$ on feature extractor;
    \STATE Calculating $\mathcal{L}_{channel}$ with discriminator $f_\phi$ following Equation~\ref{eq:diversity};
    \STATE Calculating $\mathcal{L}_{node}$ following Equation~\ref{eq:node_cls};
    \STATE Update $\theta,\xi, \phi$ with gradients to minimize Loss $\mathcal{L}_{full}$
    \ENDWHILE
    
  \RETURN Learned model parameters $\theta,\xi$
  \end{algorithmic}
\end{algorithm}

\section{Experiment}
In this section, we conduct a set of experiments to evaluate the benefits of proposed disentanglement-eliciting self-supervision signals. We apply them to the model introduced in Section ~\ref{sec:overview}, and test it on $6$ real-world graph datasets. To evaluate its flexibility, we also incorporate the designed edge disentanglement module with other typical GNN layers to report the improvements. A set of sensitivity analysis is conducted on weight of self-supervision signals, and a case study on learned edge distribution is also provided. Particularly, we want to answer the following questions:
\begin{itemize}[leftmargin=0.2in]
    \item \textbf{RQ1} Can the proposed edge disentanglement module with self-supervision signals improve the downstream task of node classification?
    \item \textbf{RQ2} Are the designed approach flexible to be applied in together with different GNN layer variants?
    \item \textbf{RQ3} Can the designed self-supervision tasks encourage edge disentanglement module to capture different relation patterns?
\end{itemize}

\subsection{Experiment Settings}

\subsubsection{Dataset}
We conduct experiments on $6$ publicly-available graph datasets, and their details are given below:
\begin{itemize}[leftmargin=*]
    \item Cora: Cora is a citation network dataset for transductive learning setting. It contains one single large graph with $2,708$ papers from $7$ areas. Each node has a $1433$-dim attribution vector, and a total number of $13,264$ citation links exist in that graph.
    \item BlogCatalog: This is a social network dataset crawled from BlogCatalog\footnote{http://www.blogcatalog.com}, with $8,652$ bloggers from $38$ classes and $501,446$ friendship edges. The dataset doesn't contain node attributes. Following ~\cite{Perozzi2014DeepWalkOL}, we attribute each node with a $64$-dim embedding vector obtained from Deepwalk.
    \item Cora\_full: A more complete version of Cora dataset. We filter out classes with less than $25$ instances, and obtain a citation network of $70$ classes with $19,793$ nodes and $146,635$ edges. 
    
    \item Squirrel and Chameleon: These two datasets~\cite{pei2019geom} are subgraphs containing web pages in Wikipedia discussing specific topics, and are typically used as graphs with certain degrees of heterophily~\cite{zhu2020beyond}. Nodes represent web pages and edges are mutual links between pages. Nodes are categorized into 5 classes based on the amount of their average monthly traffic. Squirrel contains $5,201$ nodes and $401,907$ edges, while Chameleon contains $2,277$ nodes and $65,019$ edges.
    
    \item IMDB: A movie dataset collected by ~\cite{wang2019heterogeneous}. It contains $4,780$ nodes each representing a movie, and are divided into three classes (Action, Comedy, Drama) according to their genre. In total we have $21,018$ edges.

\end{itemize}

\subsubsection{Baseline}
First, we include representative models for node classification as baselines, which include:
\begin{itemize}[leftmargin=*]
    \item MLP. A MLP with one hidden layer is applied as the feature extractor. This baseline is implemented to show the node classification accuracy when relation information is not utilized;
    \item GCN. Two GCN layers is stacked as the feature extractor. GCN~\cite{Kipf2017SemiSupervisedCW} is a popular spectral GNN variant based on graph Laplacian, and has been shown to perform well in graphs with high degree of homophily;
    \item GraphSage. Two GraphSage layers with mean aggregator are used as feature extractor. GraphSage is a spatial GNN variant introduced in ~\cite{Hamilton2017InductiveRL};
    \item GIN~\cite{xu2018powerful}. It adopts an injective multiset function as the neighborhood aggregator, and is shown to have stronger representation ability than classical GNN layers. We stack two such layers as the feature extractor;   
    \item FactorGCN~\cite{yang2020factorizable} Similar with our motivation, FactorGCN also targets at disentangling graphs with repsect to latent factors. However, it is mainly designed for graph-level tasks, and has no self-supervisions to encourage edge disentanglement. We include it as a baseline by using it as the feature extractor.
\end{itemize}

We also select two GNN variants from heterophily graph domain, which are designed to model both inter-class and intra-class edges. These models may be better in capturing rich relation information, hence are also included as baselines:
\begin{itemize}[leftmargin=*]
    \item MixHop. MixHop~\cite{abu2019mixhop} layer explicitly learns to mix feature representations of neighbors at various distances, and uses sparsity regularization to increase interpretability.
    \item H2GCN. A specially designed GNN layer proposed in ~\cite{zhu2020beyond} to work on graphs with high heterophily degrees. It enables more flexible feature aggregation from neighborhood.
\end{itemize}
To evaluate the effectiveness of proposed self-supervision signals, we also compare with the base architecture without self-supervisions, and denote it as {\method}\_Base. 

%

\subsubsection{Configurations}
All experiments are conducted on a $64$-bit machine with Nvidia GPU (Tesla V100, 1246MHz , 16 GB memory), and ADAM optimization algorithm is used to train the models. For all methods, the learning rate is initialized to $0.001$, with weight decay being $5e$-$4$. Besides, all models are trained until converging, with the maximum training epoch being $1000$.

Alternating step $n\_step$ is fixed as $5$, and $\{\lambda_1,\lambda_2,\lambda_3\}$ are set by grid search in $\{1e-4, 1e-2, 1, 10, 1e2\}$ for each dataset. Train:eval:test splits are set to $2$:$3$:$5$. Unless specified otherwise, such configurations is adopted on all datasets throughout experiments.

\subsubsection{Evaluation Metrics}
Following existing works in evaluating node classification~\cite{rout2018handling,johnson2019survey}, we adopt two criteria: classification accuracy(ACC), and macro-F score. ACC is computed on all testing examples at once, and F measure gives the harmonic mean of precision and recall for each class. We calculate F-measure separately for each class and then report their non-weighted average.

\subsection{Node Classification Performance}
To answer \textbf{RQ1}, in this section, we compare the performance on node classification between proposed {\method} and all aforementioned baselines. Models are tested on $6$ real-world datasets, and each experiment is conducted 3 times to alleviate the randomness. The average results with standard deviation are reported in Table~\ref{tab:result} and Table~\ref{tab:result2}. From the tables, we make the following observations:
\begin{itemize}
    \item Our proposed {\method} consistently outperforms baselines on all datasets with a clear margin. For example, {\method} shows an improvement of $1.71$ point in accuracy on BlogCatalog, and $1.41$ point in accuracy on IMDB compared to the best-performed baseline respectively.
    \item Comparing with {\method}\_Base in which self-supervisions are not adopted, {\method} also shows a consistent improvement. For example, in terms of accuracy, it improves for $3.67$ point on BlogCatalog and $1.72$ point on Chameleon.
\end{itemize}

To summarize, these results show the advantages of introduced edge disentanglement module and self-supervision signals in terms of node classification performance. It indicates that through automatic disentanglement and utilization of multi-relations behind edges, our approach is better at learning representations on graphs.

\begin{table*}[h!]
  \setlength{\tabcolsep}{4.5pt}
  
  \caption{Comparison of different approaches on downstream task: node classification.}\label{tab:result} 
  \vskip -1em
  \begin{tabular}{p{2.1cm} | P{1.52cm}  P{1.52cm} | P{1.52cm}  P{1.52cm} |  P{1.52cm}  P{1.52cm} | P{1.52cm}  P{1.52cm} }

    \hline
     &  \multicolumn{2}{c|}{Cora} &  \multicolumn{2}{c|}{BlogCatalog} &  \multicolumn{2}{c|}{Cora\_full} &  \multicolumn{2}{c}{IMDB} \\
    \hline
    Methods & ACC &  F Score & ACC  & F Score & ACC  & F Score & ACC  & F Score \\
    \hline 
    MLP & $62.39_{\pm0.54}$ & $59.24_{\pm0.40}$ & $28.93_{\pm0.14}$ & $20.48_{\pm0.19}$ & $43.26_{\pm0.12}$ &$32.86_{\pm0.05}$ & $50.68_{\pm0.40}$ &$41.72_{\pm0.31}$ \\
    GCN & $82.25_{\pm0.19}$ & $80.57_{\pm0.26}$ & $28.61_{\pm0.28}$ & $20.35_{\pm0.49}$ & $53.84_{\pm0.18}$ & $46.08_{\pm0.31}$ & $52.26_{\pm0.08}$ & $47.16_{\pm0.36}$\\
    GraphSage & $80.27_{\pm0.52}$ & $79.11_{\pm0.48}$ & $29.01_{\pm0.21}$ & $20.41_{\pm0.06}$ &$49.06_{\pm0.08}$& $39.47_{\pm0.32}$ & $53.31_{\pm0.06}$ & $43.38_{\pm0.14}$ \\
    GIN & $80.39_{\pm0.08}$ & $78.77_{\pm0.13}$ & $26.71_{\pm1.13}$ & $17.75_{\pm0.48}$ & $45.99_{\pm0.10}$& $38.21_{\pm0.42}$ & $52.47_{\pm0.35}$& $45.38_{\pm0.46}$ \\
    FactorGCN &  $82.14_{\pm0.18}$ & $80.39_{\pm0.11}$ & $25.35_{\pm0.08}$ & $17.40_{\pm0.10}$ & $44.08_{\pm0.32}$& $36.60_{\pm0.56}$  & $54.52_{\pm0.11}$ & $47.16_{\pm0.36}$ \\
    \hline
    MixHop & $82.07_{\pm0.17}$ & $81.07_{\pm0.56}$ & $26.18_{\pm0.69}$ & $17.68_{\pm0.94}$ & $49.19_{\pm0.16}$&$40.66_{\pm0.27}$ &$54.16_{\pm0.20}$ & $44.62_{\pm0.25}$\\
    H2GCN & $79.58_{\pm0.23}$ & $78.50_{\pm0.14}$ & $34.11_{\pm0.25}$ & $23.21_{\pm0.80}$& $55.70_{\pm0.03}$ & $48.60_{\pm0.07}$ & $54.72_{\pm0.23}$ & $44.48_{\pm0.57}$\\
    \hline
    DisGNN\_Base & $81.81_{\pm0.42}$ & $80.47_{\pm0.33}$ & $32.15_{\pm0.26}$&$23.22_{\pm0.38}$ &$57.20_{\pm0.04}$ &$50.23_{\pm0.14}$ &$54.98_{\pm0.47}$ & $47.02_{\pm0.32}$\\
    DisGNN  & $\mathbf{83.16}_{\pm0.48}$ & $\mathbf{81.99}_{\pm0.71}$ &$\mathbf{35.82}_{\pm0.19}$ & $\mathbf{23.94}_{\pm1.04}$ & $\mathbf{58.83}_{\pm0.17}$& $\mathbf{51.13}_{\pm0.34}$& $\mathbf{56.39}_{\pm0.14}$& $\mathbf{47.73}_{\pm0.34}$\\
    \hline
  \end{tabular}
\end{table*}

\begin{table}[h!]
  \setlength{\tabcolsep}{4.5pt}
  
  \caption{Comparison of different approaches on downstream task: node classification on two heterophily graphs.}\label{tab:result2} 
  \vskip -1em
  \begin{tabular}{p{1.8cm} | P{1.4cm}  P{1.4cm} | P{1.4cm}  P{1.4cm}}

    \hline
     &  \multicolumn{2}{c|}{Chameleon} &  \multicolumn{2}{c}{Squirrel} \\
    \hline
    Methods & ACC  & F Score & ACC & F Score \\
    \hline
    MLP & $35.71_{\pm0.85}$ & $34.59_{\pm1.16}$ & $25.16_{\pm0.51}$ & $25.17_{\pm0.50}$ \\
    GCN & $63.17_{\pm0.54}$& $63.04_{\pm0.54}$ & $51.69_{\pm0.18}$ & $51.50_{\pm0.05}$ \\
    GraphSage & $46.45_{\pm0.69}$ &$46.33_{\pm0.41}$ & $32.34_{\pm0.35}$ & $32.53_{\pm0.78}$ \\
    GIN & $41.50_{\pm0.38}$& $41.21_{\pm0.34}$ & $33.20_{\pm0.50}$ & $31.67_{\pm0.41}$ \\
    FactorGCN &$61.56_{\pm0.21}$ & $61.36_{\pm0.20}$& $47.07_{\pm0.50}$  & $46.80_{\pm0.56}$ \\
    \hline
    MixHop & $53.63_{\pm0.65}$&$53.40_{\pm0.64}$ & $39.72_{\pm0.54}$ &   $39.63_{\pm0.48}$ \\
    H2GCN & $63.48_{\pm0.20}$& $63.37_{\pm0.14}$ & $49.81_{\pm0.39}$ & $49.90_{\pm0.43}$  \\
    \hline
    DisGNN\_Base &$62.73_{\pm1.15}$ &$62.83_{\pm0.60}$ & $51.48_{\pm0.40}$ &  $51.59_{\pm0.43}$ \\
    DisGNN  &$\mathbf{64.45}_{\pm0.16}$ & $\mathbf{64.39}_{\pm0.16}$ & $\mathbf{52.51}_{\pm0.24}$ & $\mathbf{52.35}_{\pm0.26}$  \\
    \hline
  \end{tabular}
\end{table}

\subsection{Combination with Different GNN Layers}
To answer \textbf{RQ2}, we modify the architecture of {\method} by replacing the message aggregation and node update mechanism in Equation~\ref{eq:GNN_main} with other GNN variants. Specifially, we test on GCN and GraphSage layers due to their popularity. Experiments are randomly conducted for $3$ times on Cora, BlogCatalog, IMDB, and Squirrel. Results w/o SSL signals are both reported in Table~\ref{tab:LayerType}. 

From the result, a consistent improvement can be observed on these four dataset with both GCN and GraphSage layers. It is shown that the proposed approach is effective when incorporated into other GNN variants, validating the generality and advantage of proposed edge disentanglement.

\begin{table}[h!]
  \setlength{\tabcolsep}{4.5pt}
  \small
  \caption{Node classification accuracy when incorporating proposed edge disentanglement with different GNNs.} \label{tab:LayerType}
  \vskip -1em
  \begin{tabular}{P{1.7cm} || P{1.4cm} | P{1.4cm} | P{1.4cm} | P{1.4cm}  }
    \hline
     &  Cora & BlogCatalog  & IMDB & Squirrel \\
    \hline
    GCN\_Base & $82.21_{\pm0.36}$ & $33.74_{\pm0.21}$  & $55.23_{\pm0.48}$ &  $51.42_{\pm0.32}$ \\
    DisGNN-GCN &  $83.89_{\pm0.62}$  & $35.09_{\pm0.57}$ & $57.06_{\pm0.67}$ & $52.62_{\pm0.29}$  \\
    \hline
    Sage\_Base &  $79.61_{\pm0.06}$ & $33.16_{\pm0.35}$ & $54.21_{\pm0.19}$&$32.67_{\pm0.36}$  \\
    DisGNN-Sage &  $81.17_{\pm0.18}$ & $34.61_{\pm0.44}$ & $55.63_{\pm0.27}$ &$34.49_{\pm0.27}$  \\
    
    \hline
  \end{tabular}
\end{table}

\subsection{Comparison with Different SSL Signals}
To further evaluate the effectiveness of our proposed SSL signals, we compare it with three other typical self-supervision tasks: Masked attribute prediction~\cite{hu2019strategies}, PairwiseDistance~\cite{jin2020self}, and Context prediction~\cite{hu2019strategies}.
These tasks are implemented on the {\method}\_Base, with the same optimization and configuration as {\method}. Their weights are found via grid search in $\{1e-4, 1e-2, 1, 10, 1e2, 1e3\}$. 

Comparisons are summarized in Table~\ref{tab:SSL}. We can observe that {\method} is the most effective approach across these settings. This result validates our proposal of utilizing latent relations behind edges to improve the learning on graphs. Note that {\method} can also be utilized in together with these SSL tasks, but we leave that for the future as it is not the focus of this work.

\begin{table*}[h]
  \setlength{\tabcolsep}{4.5pt}
  
  \caption{Comparison with different SSL signals.}\label{tab:SSL} 
  \vskip -1em
  \begin{tabular}{p{2.1cm} | P{1.52cm}  P{1.52cm} | P{1.52cm}  P{1.52cm} |  P{1.52cm}  P{1.52cm} | P{1.52cm}  P{1.52cm} }

    \hline
     &  \multicolumn{2}{c|}{Cora} &  \multicolumn{2}{c|}{BlogCatalog} &  \multicolumn{2}{c|}{IMDB}  &  \multicolumn{2}{c}{Chameleon}\\
    \hline
    Methods & ACC &  F Score & ACC  & F Score & ACC  & F Score & ACC  & F Score \\

    \hline
    DisGNN\_Base & $81.81_{\pm0.42}$ & $80.47_{\pm0.33}$ & $32.15_{\pm0.26}$&$23.22_{\pm0.38}$  &$54.98_{\pm0.47}$ & $47.02_{\pm0.32}$ & $62.73_{\pm1.15}$ &$62.83_{\pm0.60}$ \\
    +MaskedAttr  & $82.70_{\pm0.45}$ & $81.41_{\pm0.25}$ & $34.47_{\pm0.21}$ & $23.52_{\pm0.69}$ & $53.54_{\pm0.24}$ & $47.27_{\pm0.40}$ &$64.23_{\pm0.53}$ & $64.20_{\pm0.54}$ \\
    +DistancePred  & $82.74_{\pm0.36}$ & $81.37_{\pm0.28}$  & $34.72_{\pm0.16}$  & $23.53_{\pm0.51}$  &$56.07_{\pm0.24}$ &$47.03_{\pm0.59}$ & $63.87_{\pm0.25}$ & $63.63_{\pm0.29}$ \\
    +ContextPred  & $82.81_{\pm0.57}$ & $81.52_{\pm0.57}$ & $35.13_{\pm0.19}$ & $23.43_{\pm0.31}$ & $55.21_{\pm0.51}$& $46.43_{\pm0.50}$  & $62.05_{\pm0.36}$ & $61.83_{\pm0.53}$ \\
    
    DisGNN  & $\mathbf{83.16}_{\pm0.48}$ & $\mathbf{81.99}_{\pm0.71}$ &$\mathbf{35.82}_{\pm0.19}$ & $\mathbf{23.94}_{\pm1.04}$ &  $\mathbf{56.39}_{\pm0.14}$& $\mathbf{47.73}_{\pm0.34}$ & $\mathbf{64.45}_{\pm0.16}$ & $\mathbf{64.39}_{\pm0.16}$  \\
    \hline
  \end{tabular}
\end{table*}

\subsection{Ablation Study}
In this section, we conduct a series of ablation studies and sensitivity analysis to evaluate the influence of each component in {\method}. 

\subsubsection{Edge Disentanglement Module Design}

\begin{figure}[t]
  \centering
		\includegraphics[width=0.45\textwidth]{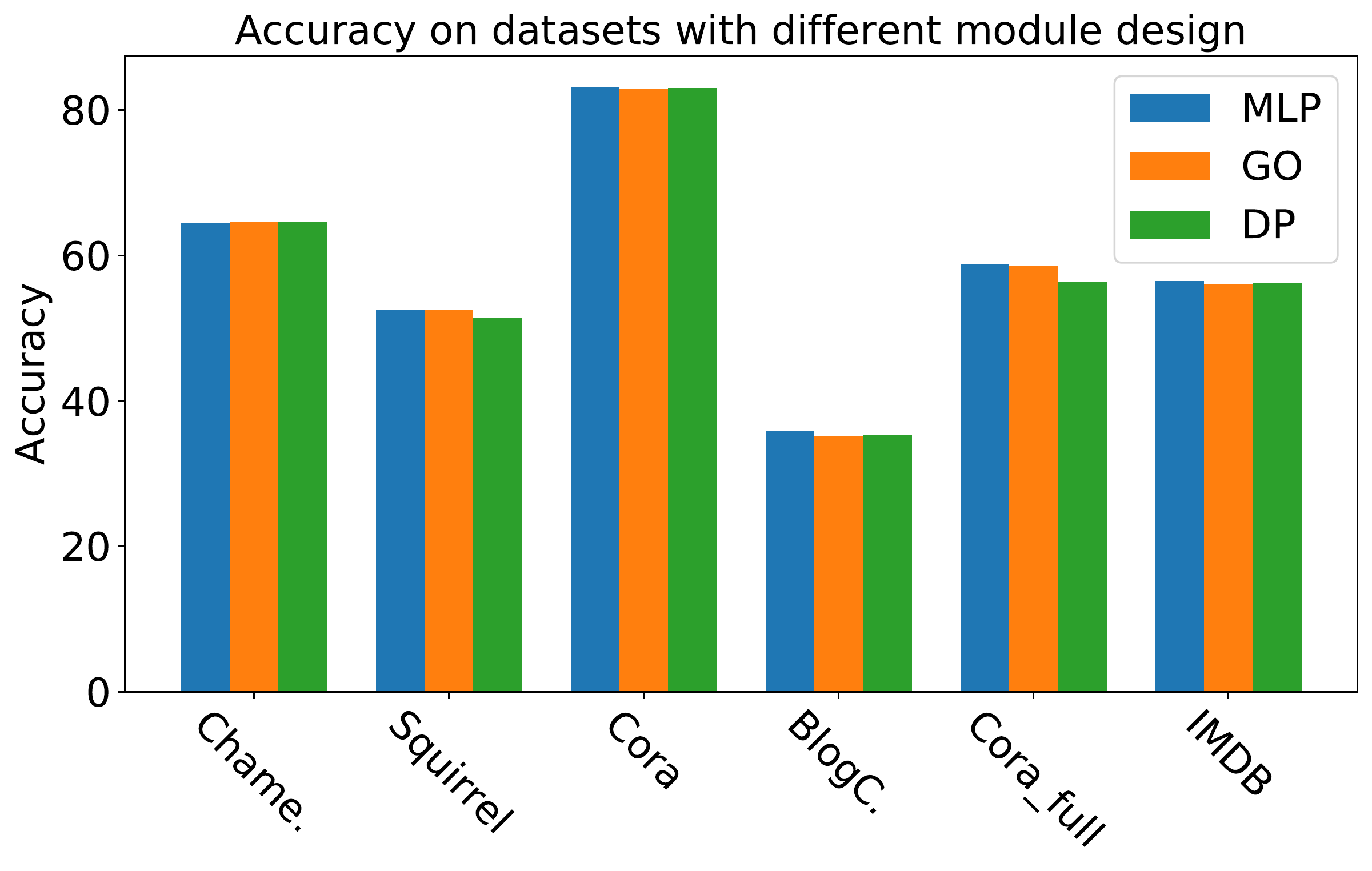}
		
    \vskip -1em
    \caption{Influence of edge disentanglement module design.} \label{fig:disentangle_module}
  \setlength{\abovecaptionskip}{0cm}
\end{figure}

Different choices of edge-modeling mechanism could influence the representation ability of it in capturing relations~\cite{kim2020find}. To evaluate its significance, we tested two other widely-used designs in replace of the MLP-based disentanglement module used in Equation~\ref{eq:used_at}:
\begin{itemize}[leftmargin=*]
    \item Single-layer neural network (GO): uses a linear layer on the concatenation of node pair representations:
    \begin{equation}
        \hat{e}_{v,u}^{l,i} = sigmoid \big( (\mathbf{a}^{l,i})^T  \big([ \mathbf{W}_{1}^{l,i} \mathbf{h}^{l}_{v} || \mathbf{W}_{1}^{l,i}\mathbf{h}^{l}_{u} ] \big) \big)
    \end{equation}
    
    \item Dot-product (DP): uses inner-production on node embeddings to obtain pair-wise similarity:
    \begin{equation}
    \hat{e}_{v,u}^{l,i} = sigmoid \big( \big(\mathbf{W}_{1}^{l,i} \mathbf{h}^{l}_{v}\big)^T \cdot \mathbf{W}_{1}^{l,i} \mathbf{h}^{l}_{v} \big) \big) 
    \end{equation}
\end{itemize}
Experiments are randomly conducted for $3$ times, and all other configurations remained unchanged. The results are summarized in Figure~\ref{fig:disentangle_module}. From the result, we can observe that in most cases, MLP-based disentanglement performs best while DP-based disentanglement performs worst. Generally, the performance gap is clearer on large graphs like Cora\_full, and smaller on small grpahs like Chameleon. We attribute this phenomenon to the greater expressive capacity of MLP over GO and DP.


\subsubsection{SSL Signal Weight}
\vskip -2em
\begin{figure}[h]
  \centering
    \subfigure[Weight of edge recovery loss ]{
		\includegraphics[width=0.23\textwidth]{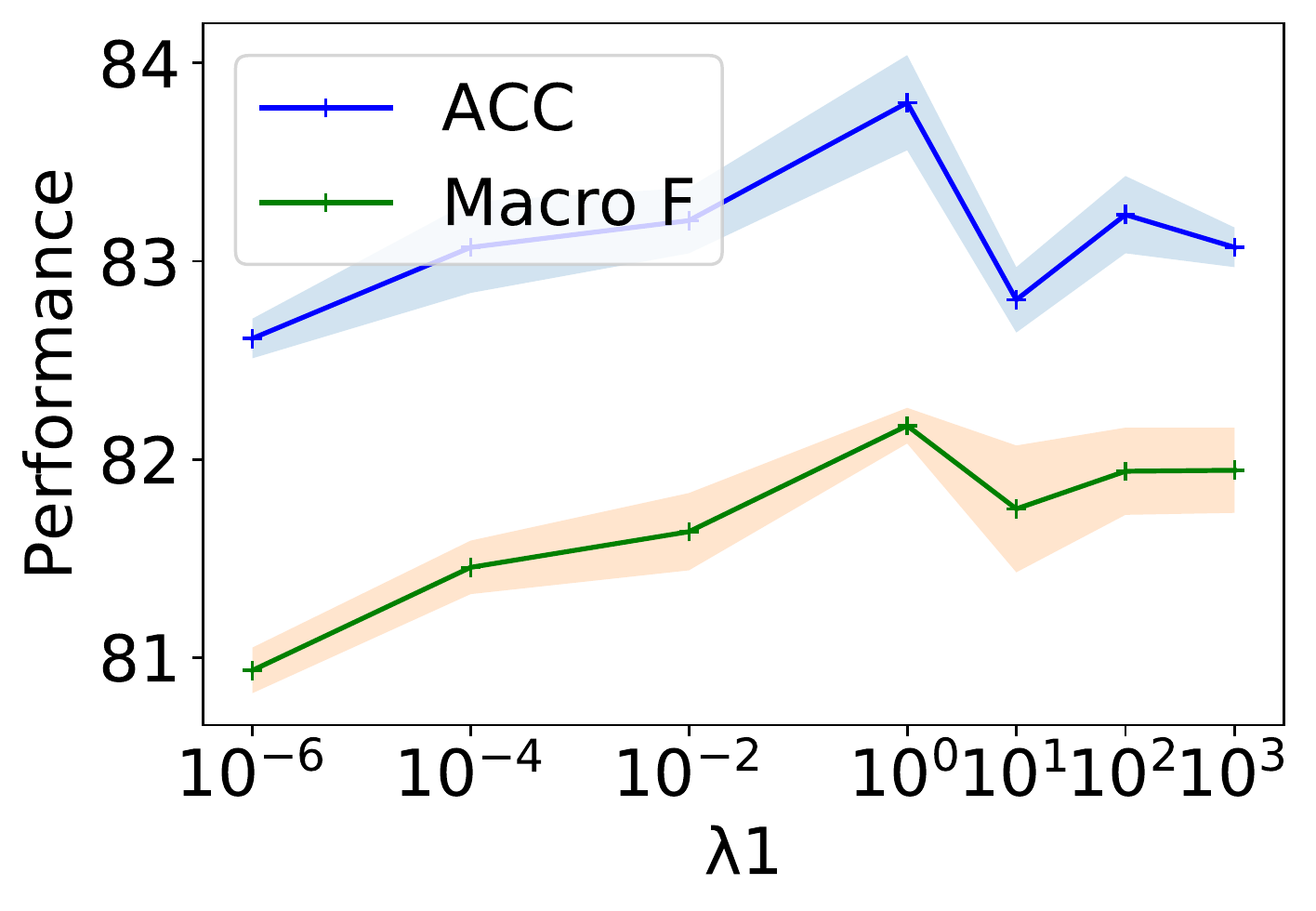}}
    \subfigure[Weight of label conformity loss ]{
		\includegraphics[width=0.23\textwidth]{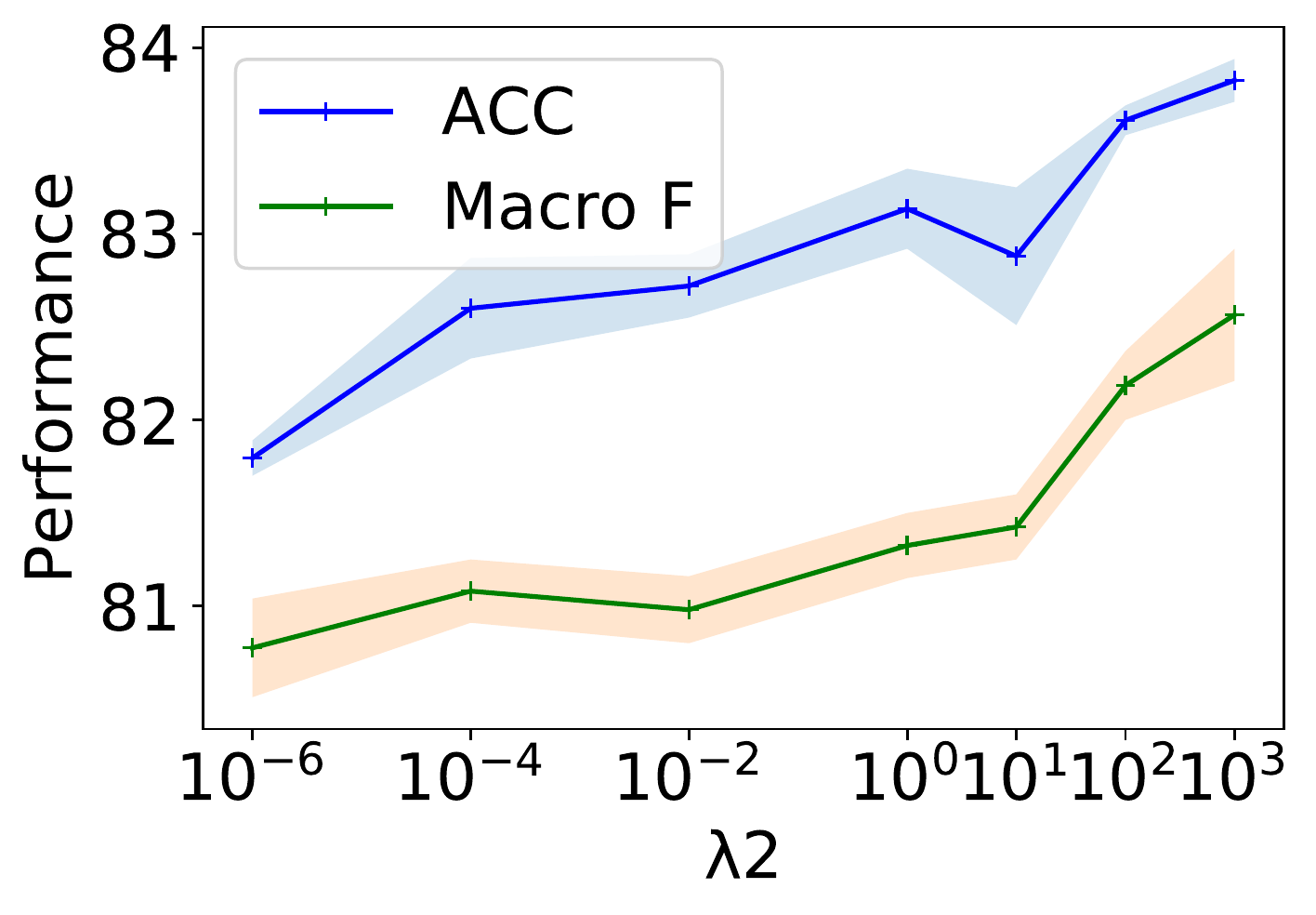}}
    \subfigure[Weight of channel difference loss ]{
		\includegraphics[width=0.23\textwidth]{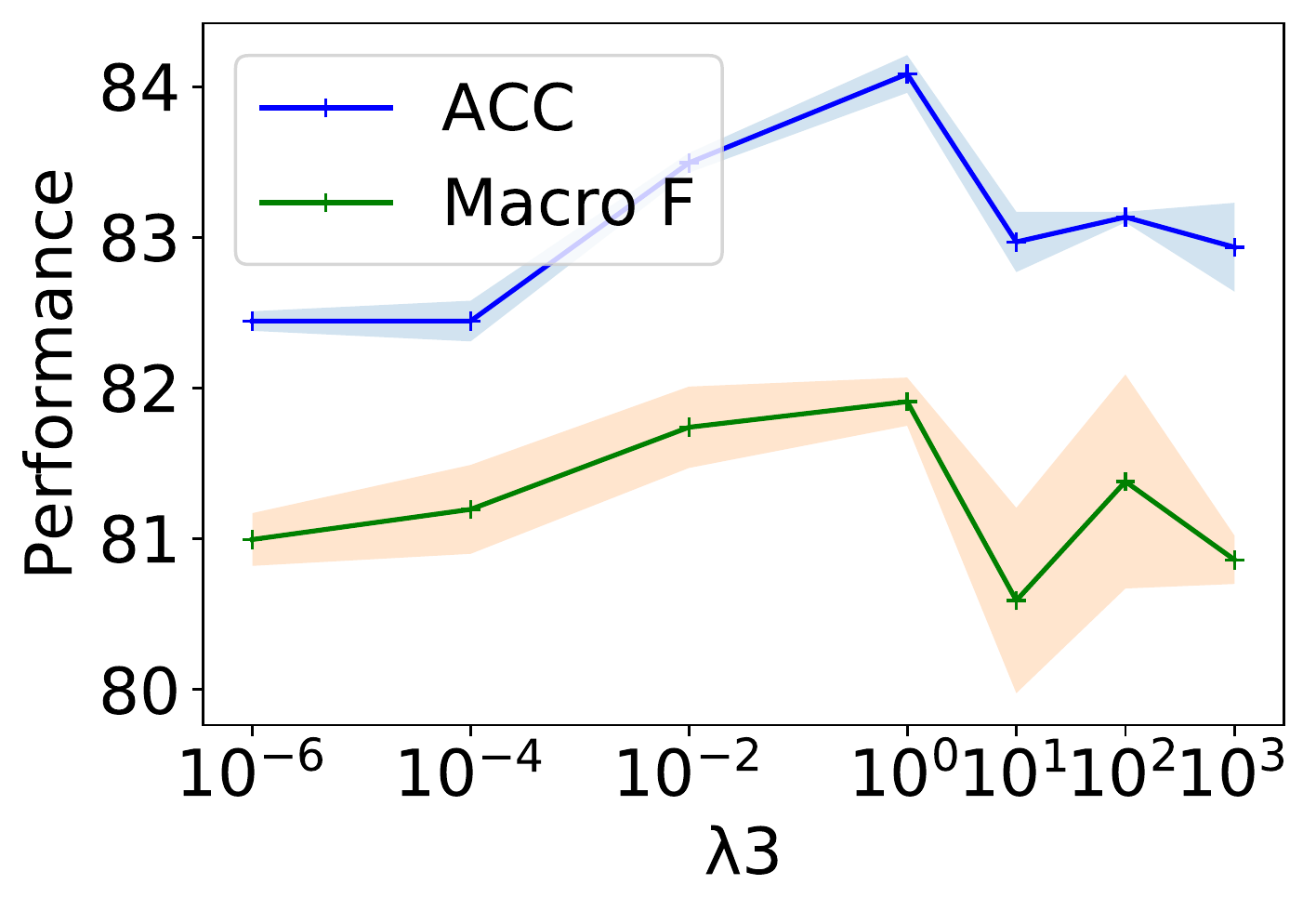}}
	\vskip -1.5em
   \caption{Parameter Sensitivity on Cora. $\lambda_1, \lambda_2, \lambda_3 $ control the weight of each SSL signal respectively. }\label{fig:parameter_sensitivity}
\end{figure}


In this part, we vary the hyper-parameter $\lambda_1$, $\lambda_2$, $\lambda_3$ to test {\method}'s sensitivity towards them. These three values control the weights of each proposed SSL signal respectively. To keep simplicity, we keep all other configurations unchanged, and adopt only one SSL task at a time. Its corresponding $\lambda$ varies in $\{1e-6,1e-4, 1e-2, 1, 10,1e2, 1e3\}$, and experiments are randomly conducted for $3$ times on Cora.

From the result shown in Figure~\ref{fig:parameter_sensitivity}, we can observe that on Cora, increasing weight of Edge Recovery or Channel Difference is beneficial within the range $[0,1]$, and further increasing that would result in performance drop. While for Label Conformity, its benefits is clear with a larger weight like with the range $[1e2,1e3]$.

\subsection{Case Study}

\begin{figure}[t]
  \centering
  \subfigure[DisGNN\_Base]{
		\includegraphics[width=0.23\textwidth]{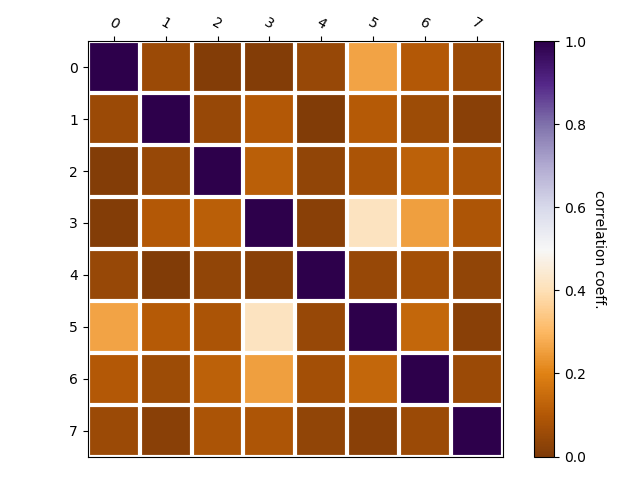}}
    \subfigure[DisGNN]{
		\includegraphics[width=0.23\textwidth]{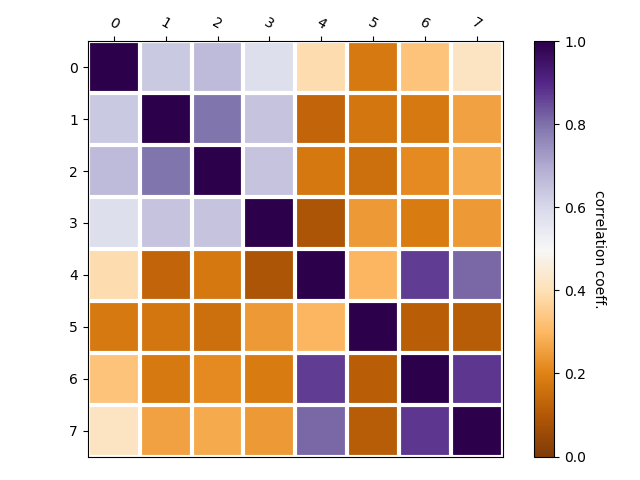}\label{fig:caseb}}
		
    \vskip -1em
    \caption{Correlation coefficient among edges captured by different channels on Cora, obtained from the first layer. } \label{fig:case}
  \setlength{\abovecaptionskip}{0cm}
\end{figure}
In most cases, there is no ground-truth for disentangled factors behind edges in real-world graphs, making the direct evaluation of edge disentanglement difficult. To provide insights into the behavior of {\method} and partly answer \textbf{RQ3}, we conduct a case study and compare the distribution of disentangled edges w/o proposed self-supervision tasks. Concretely, Pearson correlation coefficients among edges captured by each channel are shown in Figure~\ref{fig:case}.

Through comparing with {\method}\_base, we can see that when proposed SSL tasks are used, a clear grid-like structure can be observed from correlation among channels. In Figure~\ref{fig:caseb}, the top four channels and the bottom four channels have little correlation as they focus on homo-edges and hetero-edges respectively. Within these two groups, we can further observe that channel $1,2$ hold a higher correlation with each other, similar is the case of channels $4,6$, and $7$. While channel $5$ has a low correlation towards other channels within the same group. This result shows that although they all model hetero-edges, they still capture distinct patterns.

\section{Conclusion}
In this work, we studies the discovering and utilizing of latent relations behind edges to facilitate node classification, one typical task on learning from graphs. An edge disentanglement module is designed and incorporated into modern GNN layers, and three disentanglement-eliciting SSL signals are proposed to optimize jointly with node classification task. Experiments validate the advantage of the proposed {\method}, showing the benefits of exploiting latent multi-relation nature of edges.

In the future, works can be done to discover more heuristics promoting the automatic edge disentanglement. Due to the lack of explicit supervision, it is remains an open problem. Furthermore, the benefits of disentanglement towards other graph learning tasks, like edge-level or graph-level tasks, is also a promising direction.

\section{Acknowledgement}
This material is based upon work supported by, or in part by, the National Science Foundation under grants number IIS-1707548, CBET-1638320, IIS-1909702, IIS-1955851, and the Army Research Office under grant W911NF21-1-0198. The findings and conclusions in this paper do not necessarily reflect the view of the funding agency.

\bibliographystyle{ACM-Reference-Format}
\bibliography{acmart}

\end{document}